\documentstyle[12pt,prd,graphics,epsfig,aps,]{revtex}
\begin{document}
\title{New results of $^{116}$Cd double $\beta $ decay study with $^{116}$CdWO$_4$
scintillators}
\author{F.A.~Danevich, A.Sh.~Georgadze, V.V.~Kobychev, B.N.~Kropivyansky,
A.S.~Nikolaiko, O.A.~Ponkratenko, V.I.~Tretyak, S.Yu.~Zdesenko,
Yu.G.~Zdesenko\footnote{Corresponding author: zdesenko@kinr.kiev.ua}}
\address{Institute for Nuclear Research, MSP 03680 Kiev, Ukraine}
\author{P.G.~Bizzeti, T.F.~Fazzini, P.R.~Maurenzig}
\address{Dip. di Fisica, Universit\'a di Firenze and INFN, 50125 Firenze, Italy}
\date{\today }
\maketitle

\begin{abstract}
{
A new phase of a $^{116}$Cd double $\beta $ decay experiment is in progress
in the Solotvina Underground Laboratory. Four enriched $^{116}$CdWO$_4$
scintillators with total mass of 339 g are used in a set up, whose active
shield is made of 15 natural CdWO$_4$ crystals (20.6 kg). The background
rate in the energy interval 2.5--3.2 MeV is 0.03 counts/yr$\cdot $kg$\cdot $%
keV. The half-life for 2$\nu $2$\beta $ decay of $^{116}$Cd is measured as $%
T_{1/2}(2\nu )=2.6\pm 0.1$(stat)$_{-0.4}^{+0.7}$(syst)$\times 10^{19}$ yr.
The $T_{1/2}$ limits for neutrinoless 2$\beta $ decay of $^{116}$Cd are set
at $T_{1/2}$ $\geq 0.7(2.5$)}$\times ${1$0^{23}$ yr at 90\%(68\%) C.L. for
transition to ground state of $^{116}$Sn, while for decays to the first 2$%
_1^{+}$ and second 0$_1^{+}$ excited levels of $^{116}$Sn at $T_{1/2}$ $\geq
1.3(4.8$)}$\times ${1$0^{22}$ yr and $\geq 0.7(2.4$)}$\times ${1$0^{22}$ yr
with 90\%(68\%) C.L., respectively. For $0\nu 2\beta $ decay with emission
of one or two Majorons, the limits are $T_{1/2}$($0\nu $M1) $\geq 3.7(5.8$)}$%
\times ${1$0^{21}$ yr and $T_{1/2}$($0\nu $M2) $\geq 5.9(9.4$)}$\times ${1$%
0^{20}$ yr at 90\%(68\%) C.L. Restrictions on the value of the neutrino
mass, right-handed admixtures in the weak interaction, and the
neutrino-Majoron coupling constant are derived as: $m_\nu \leq 2.6(1.4)$ eV, 
$\eta \leq 3.$9}$\times ${1$0^{-8}$, $\lambda \leq 3.$4}$\times ${1$0^{-6}$,
and $g_M\leq 12(9.5$)}$\times ${1$0^{-5}$ at 90\%(68\%) C.L., respectively. }
\end{abstract}

\draft



\section{Introduction}

Neutrinoless (0$\nu $) double $\beta $ decay is forbidden in the Standard
Model (SM) since it violates lepton number ($L$) conservation. However many
extensions of the SM incorporate $L$ violating interactions and thus could
lead to the 0$\nu $2$\beta $ decay \cite{Moe94,Theo98}. Currently, besides
conventional neutrino ($\nu $) exchange mechanism, there are many other
possibilities to trigger this process \cite{Theo98}. In that sense
neutrinoless 2$\beta $ decay has a great conceptual importance due to the
strong Schechter-Valle theorem \cite{Val82} obtained in a gauge theory of
the weak interaction that a non-vanishing 0$\nu $2$\beta $ decay rate
requires neutrino to be massive Majorana particle, independent of which
mechanism induces it. Therefore, at present 0$\nu $2$\beta $ decay is
considered as a very powerful test of new physical effects beyond the SM,
and even the absence of this process -- at the present level of sensitivity
-- would help to restrict or narrow this wide choice of theoretical models.
At the same time 0$\nu $2$\beta $ decay is very important in the light of
the measured deficit of the atmospheric neutrinos flux \cite{Neut98,SK98}
and the result of the LSND accelerator experiment \cite{Neut98,LSND95},
which could be explained by means of the neutrino oscillations requiring in
turn nonzero neutrino masses ($m_\nu $). However oscillation experiments are
sensitive to neutrino mass difference, while measured 0$\nu $2$\beta $ decay
rate can give the absolute value of the $m_\nu $ scale, and hence provide a
crucial test of $m_\nu $ models.

Despite of numerous attempts to observe 0$\nu $2$\beta $ decay from 1948 up
to present days \cite{Moe94} this process still remains unobserved. The
highest $T_{1/2}$(0$\nu $) limits were set in direct experiments with
several nuclides: $T_{1/2}\geq 10^{22}$ yr for $^{82}$Se \cite{Se82}, $%
^{100} $Mo \cite{Mo100}, $^{116}$Cd \cite{Dan99}; $T_{1/2}\geq 10^{23}$ yr
for $^{130}$Te \cite{Te130} and $^{136}$Xe \cite{Xe136}; and $T_{1/2}\geq
10^{25}$ yr for $^{76}$Ge \cite{Ge76,IGEX}.

With the aim to enlarge the number of 2$\beta $ decay candidate nuclides
studied at a sensitivity comparable with that for $^{76}$Ge and $^{136}$Xe
(neutrino mass limit of 0.5--2 eV), cadmium tungstate crystal scintillators,
enriched in $^{116}$Cd to 83\%, were developed and exploited in $^{116}$Cd
research \cite{Dan95,Dan99}. The measurements were carried out in the
Solotvina Underground Laboratory in a salt mine 430 m underground ($\simeq $%
1000 m w. e.) \cite{Zde87}. In the first phase of the experiment only one $%
^{116}$CdWO$_4$ crystal (15.2 cm$^3$) was used. It was viewed by a
photomultiplier tube (PMT) through a light-guide 51 cm long and placed
inside a plastic scintillator ($\oslash $38$\times $115 cm) which served as
veto counter. A passive shield of high purity (HP) copper (5 cm), lead (23
cm) and polyethylene (16 cm) surrounded the plastic counter. The background
rate in the energy range 2.7--2.9 MeV ($Q_{2\beta }$=2805 keV \cite{Aud95})
was equal $\approx $0.6 counts/yr$\cdot $kg$\cdot $keV. With 19175 h
statistics the half-life limit for 0$\nu $2$\beta $ decay of $^{116}$Cd was
set as $T_{1/2}$(0$\nu $) $\geq $ 3.2$\times $1$0^{22}$ yr (90\% C.L.) \cite
{Dan99}, which corresponds to the restriction on the neutrino mass $m_\nu
\leq 3.9$ eV \cite{Sta90}. Limits on 0$\nu $2$\beta $ decay with emission of
one (M1) or two (M2) Majorons were obtained too: $T_{1/2}$(0$\nu $M1) $\geq $
1.2$\times $1$0^{21}$ yr and $T_{1/2}$(0$\nu $M2) $\geq $ 2.6$\times $1$%
0^{20}$ yr (90\% C.L.) \cite{Dan98}.

In the present paper new and advanced results of $^{116}$Cd research
obtained with the help of an upgraded apparatus are described.

\section{New set-up with four $^{116}$CdWO$_4$ detectors}

\subsection{Set-up and measurements}

In order to enhance the sensitivity of the $^{116}$Cd experiment, the
following improvements were scheduled: increase of the number of $^{116}$Cd
nuclei, reduction of the background and improvement of the data taking and
processing \cite{Dan99}. With this aim the new set-up with four enriched $%
^{116}$CdWO$_4$ crystals (total mass 339 g) has been mounted in the
Solotvina Laboratory in August 1998. All materials used in the installation
were previously tested and selected for low radioactive impurities in order
to reduce their contributions to background.

\begin{figure}[tbp]
\centering
\includegraphics[width=13.cm,clip,bb=150 380 480 800] {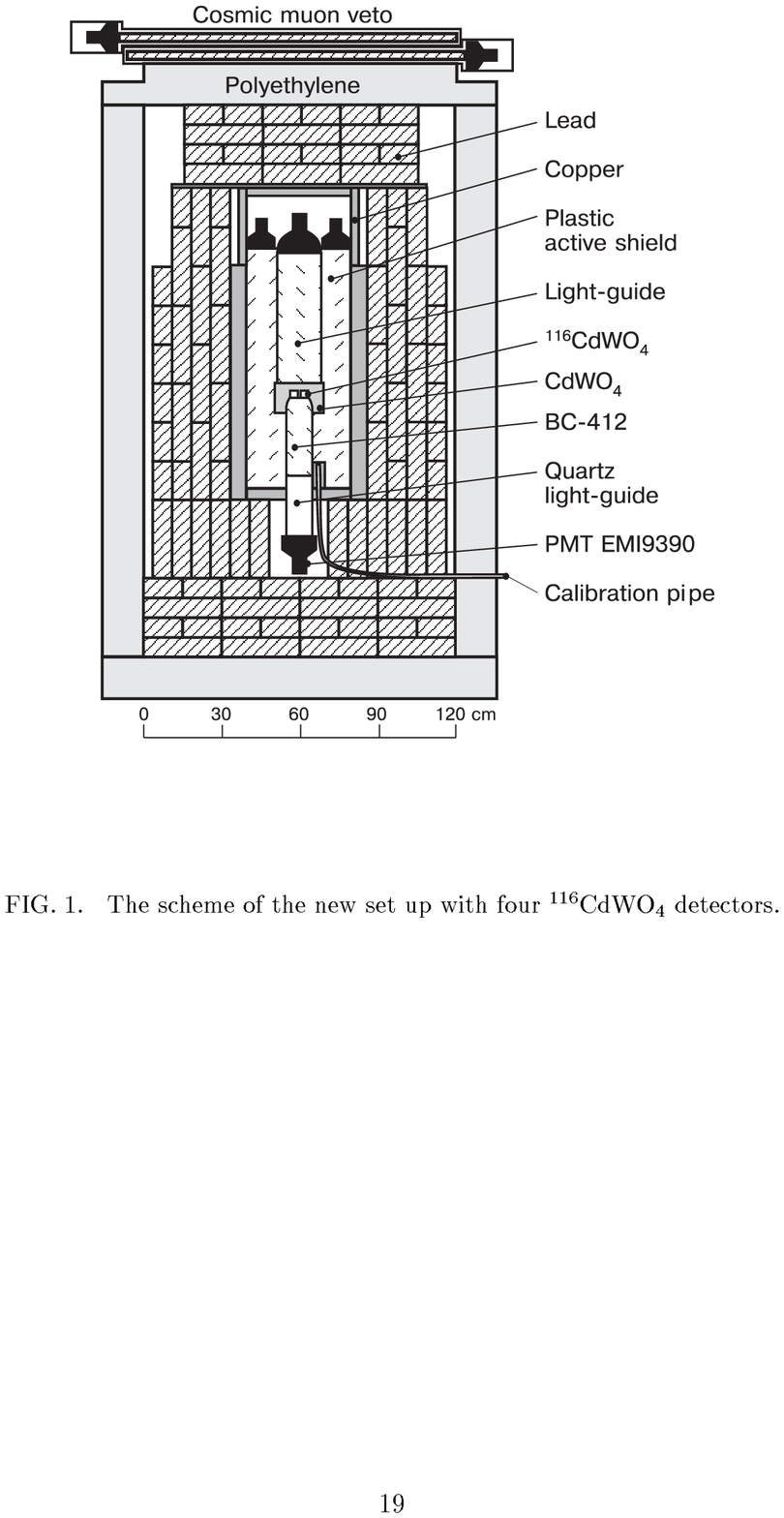}
\caption{ The scheme of the new set up with four $^{116}$CdWO$_4$ detectors.}
\label{Fig1}
\end{figure}

In the new apparatus, a scheme of which is shown in fig.~\ref{Fig1}, four
enriched crystals are viewed by the PMT (EMI9390) through one light-guide 10
cm in diameter and 55 cm long, which is composed of two glued parts: quartz
25 cm long and plastic scintillator (Bicron BC-412) 30 cm long. The $^{116}$%
CdWO$_4$ crystals are surrounded by an active shield made of 15 natural CdWO$%
_4$ scintillators of large volume ($\simeq $200 cm$^3$ each) with total mass
of 20.6 kg. Due to the high purity of the CdWO$_4$ crystals \cite{Geo96} and
their large density ($\approx $8 g/cm$^3$) this active shield reduces
background effectively. The veto crystals are viewed -- by a low background
PMT ($\oslash $17 cm) -- through an active plastic light-guide ($\oslash $17$%
\times $49 cm). In turn the whole array of CdWO$_4$ counters is placed
inside an additional active shield made of polystyrene-based plastic
scintillator with dimensions 40$\times $40$\times $95 cm. Thus, together
with both active light-guides (connected with enriched and natural crystals
on opposite sides), a complete 4$\pi $ active shield of the main $^{116}$CdWO%
$_4$ detectors is provided.

The outer passive shield consists of HP copper (thickness $3$--$6$ cm), lead
($22.5$--$30$ cm) and polyethylene (16 cm). Two plastic scintillators (120$%
\times $130$\times $3 cm) are installed above the passive shield to provide
a cosmic muons veto. Because air in the Solotvina Laboratory can be
contaminated by radon (at the level $\leq $ 30 Bq/m$^3$) the set-up is
isolated carefully against air penetration. All cavities inside the shield
are filled up by pieces of plexiglass, and HP Cu shield is sealed with the
help of silicon glue and enclosed inside a tight mylar envelope.

The new event-by-event data acquisition is based on two IBM personal
computers (PC) and a CAMAC crate with electronic units. For each event the
following information is stored on the hard disc of the first computer: the
amplitude (energy) of a signal, its arrival time and the following
additional tags: the coincidence between different detectors; the signal of
radio-noise detection system; triggers for light emitting diode (LED) and
pulse shape digitizer. The second computer records the pulse shape of the $%
^{116}$CdWO$_4$ scintillators in the energy range 0.25--5 MeV. This
complementary system is developed on the basis of a fast 12 bit ADC (Analog
Devices AD9022) and is connected with computer by parallel digital I/O board
(PC-DIO-24 from National Instruments) \cite{Faz98}. Two PC-DIO-24 boards are
used to link both computers and establish -- with the help of proper
software -- a one-to-one correspondence between the pulse shape data
recorded by the second computer and the information stored in the first PC.

The energy scale and resolution of the main detector -- four enriched
crystals taken as a whole -- were determined in the measurements with
different sources ($^{22}$Na, $^{40}$K, $^{60}$Co, $^{137}$Cs, $^{207}$Bi, $%
^{226}$Ra, $^{232}$Th and $^{241}$Am). The energy dependence of the
resolution can be expressed (for the energy above 50 keV) as $FWHM($keV) =$%
\sqrt{-226+\text{1}6.6E+6.42\times 10^{-3}E^2}$, where energy $E$ is in keV.
For instance, the resolution ($FWHM$) was equal to 14.5\% at 1064 keV and
11\% at 2615 keV. The full energy peaks are well fitted in the energy region
0.06--2.6 MeV by a Gaussian function with typical value $\chi ^2$ =
0.8--1.9. Moreover, the calibration spectra of the $^{232}$Th source were
simulated with the help of GEANT3.21 package \cite{GEANT} and event
generator DECAY4 \cite{Decay4} (the last defines initial kinematics of the
events). The simulated $^{232}$Th spectra are in good agreement with the
measured ones confirming the assumption of a Gaussian peak shape. In
particular for the 2615 keV peak of $^{208}$Tl -- which is close to the 2$%
\beta $ decay energy of $^{116}$Cd -- the value of $\chi ^2$ =1.3.

Also, the relative light yield for for $\alpha $ particles as compared with
that for electrons ($\alpha /\beta $ ratio) and energy resolution were
measured with $\alpha $ sources and corrected by using time-amplitude
analysis (see below) as following: $\alpha /\beta $ = 0.15(1) $+$7$\times $1$%
0^{-6}E_\alpha $ and $FWHM_\alpha ($keV) = $0.053E_\alpha $ ($E_\alpha $ is
in keV). The routine calibration is carried out weekly with a $^{207}$Bi
source (570, 1064 and 1770 keV) and once per two weeks with $^{232}$Th (2615
keV). The dead time of the spectrometer and data acquisition is monitored
permanently with the help of an LED optically connected with the main PMT.
The actual dead time value is $\approx $4.2$\%$ ($\approx $3\% is owing to
random coincidence between the main and shield detectors; $\approx $1.2\% is
caused by miscounts of the data acquisition).

\begin{figure}[tbp]
\centering
\includegraphics[width=18.cm,clip,bb=0 150 590 670] {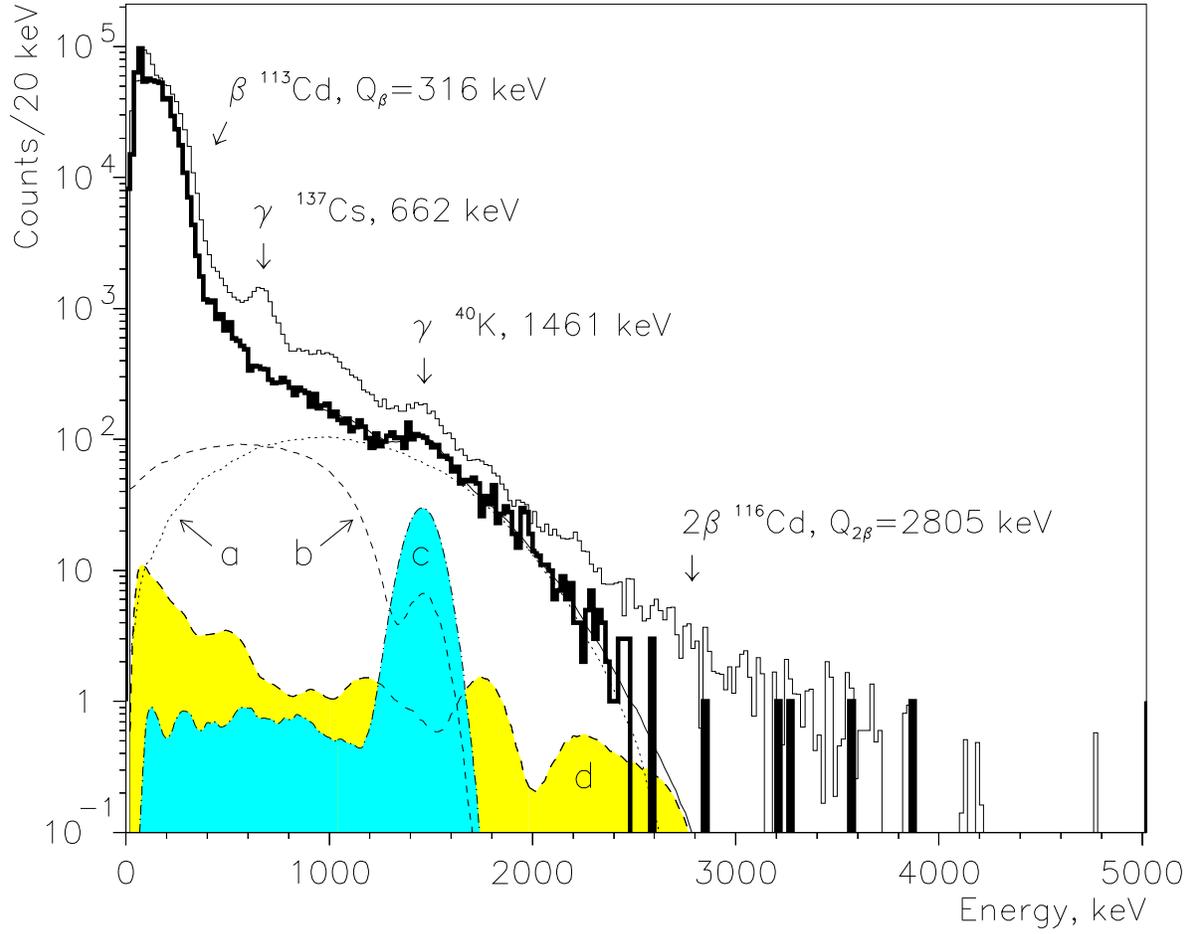}
\caption{ Background spectrum of $^{116}$CdWO$_4$ detectors (339 g) measured
in the set up with four enriched crystals during 4629 h (solid histogram).
The old data obtained with one $^{116}$CdWO$_4$ crystal (121 g; 19986 h) is
shown for comparison (thin histogram; the data are normalized to 4629 h and
mass of the new detector). The background components used for ô fit in the
energy region $900-2900$ keV: (a) 2$\nu $2$\beta $ decay of $^{116}$Cd (fit
value is $T_{1/2}$(2$\nu )=$ 2.6(1)$\times $1$0^{19}$ yr); (b) $^{40}$K inside
the $^{116}$CdWO$_4$ detector (activity value from the fit is 0.8(2)
mBq/kg); (c) $^{40}$K in the shielding CdWO$_4$ crystals (fit value is
2.1(3) mBq/kg); (d) $^{226}$Ra and $^{232}$Th contamination of PMTs.}
\label{Fig2}
\end{figure}

The background spectrum measured during 4629 h in the new installation with
four $^{116}$CdWO$_4$ crystals is given in fig.~\ref{Fig2}, where the old
data obtained with one $^{116}$CdWO$_4$ crystal of 121 g are also shown for
comparison. As it is visible from this figure, the background is decreased
in the whole energy range, except for the $\beta $ spectrum of $^{113}$Cd ($%
Q_\beta $ = $316$ keV), whose abundance in $^{116}$CdWO$_4$ crystals is $%
\approx $2\% \cite{Dan95}. In the energy region $2.5$--$3.2$ MeV -- where
the peak of 0$\nu $2$\beta $ decay of $^{116}$Cd is expected -- the
background rate is reduced to a value of 0.03 counts/yr$\cdot $kg$\cdot $keV
(only 4 events in the energy window $2.5$--$3.2$ MeV were detected during
4629 h), twenty times lower than in the previous set-up. It is achieved,
first, due to improvement of passive and active shield, and secondly, as a
result of data processing advance (time-amplitude and pulse-shape analysis),
which are described below.

\subsection{Time-amplitude analysis of the data}

\begin{figure}[tbp]
\centering
\includegraphics[width=15.cm,clip,bb=0 80 590 700] {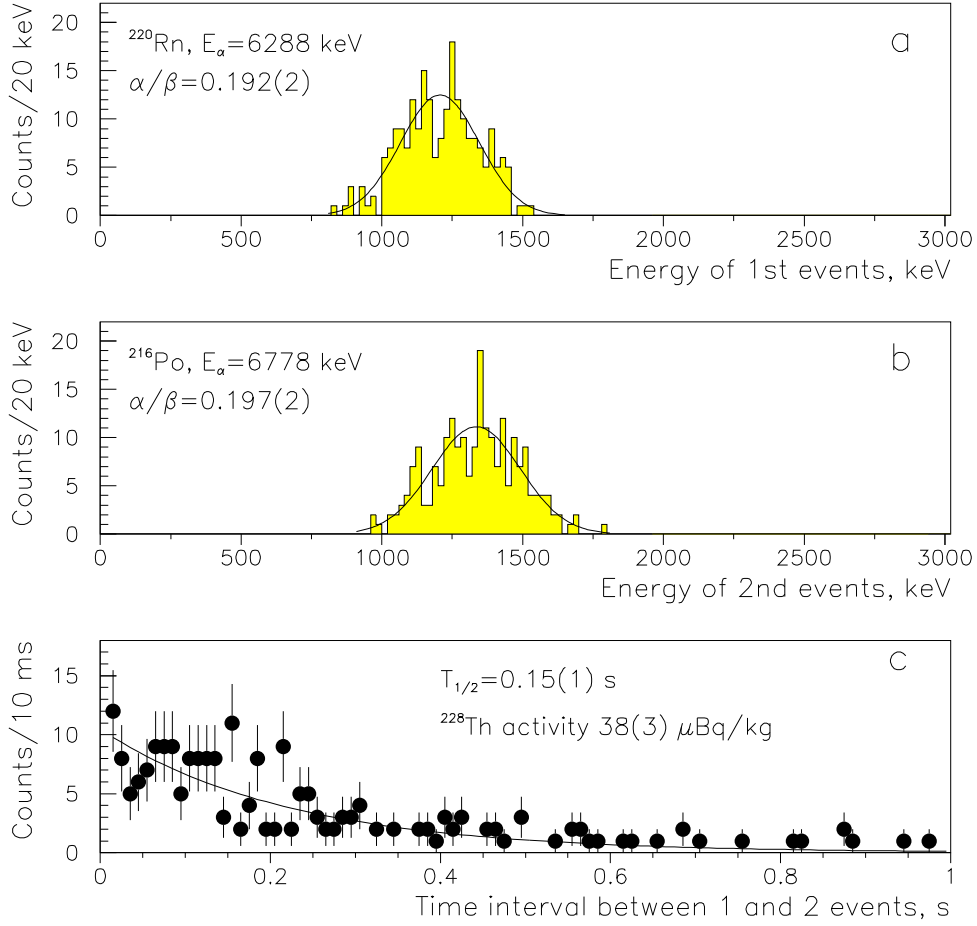}
\caption{ The energy spectra of the first (a) and second (b) $\alpha $
particles from the $^{220}$Rn $\rightarrow $ $^{216}$Po $\rightarrow $ $%
^{212}$Pb chain selected by time-amplitude analysis from $^{116}$CdWO$_4$
data. Their equivalent energies in the $\beta /\gamma $ energy scale are
near 5 times smaller because the relative light yield for $\alpha $
particles as compared with that for electrons ($\alpha /\beta $ ratio) is $%
\approx 0.2$. (c) Time distribution between the first and second events
together with exponential fit ($T_{1/2}$ = $0.15(1)$ s, while the table
value is $T_{1/2}$ = $0.145(2)$ s [24]).}
\label{Fig3}
\end{figure}

The energy and arrival time of each event can be used for analysis and
selection of some decay chains in $^{232}$Th, $^{235}$U and $^{238}$U
families (see f. e. ref. \cite{Xe136,Dan95}). As an example (important in
the following for the background rejection in the energy range of 0$\nu
2\beta $ decay), we consider here in detail the time-amplitude analysis of
the following sequence of $\alpha $ decays from $^{232}$Th family: $^{220}$%
Rn ($Q_\alpha $ = $6.40$ MeV, $T_{1/2}$ = $55.6$ s) $\rightarrow $ $^{216}$%
Po ($Q_\alpha $ = $6.91$ MeV, $T_{1/2}$ = $0.145$ s) $\rightarrow $ $^{212}$%
Pb. Because the energy of $\alpha $ particles from $^{220}$Rn decay
corresponds to $\simeq $1.2 MeV in $\beta /\gamma $ scale of $^{116}$CdWO$_4$
detector, the events in the energy region $0.7$--$1.8$ MeV were used as
triggers. Then all events (within $0.9$--$1.9$ MeV) following the triggers
in the time interval $10$--$1000$ ms (containing 94.5\% of $^{216}$Po
decays) were selected. The spectra of the $^{220}$Rn and $^{216}$Po $\alpha $
decays obtained in this way from data -- as well as the distribution of the
time intervals between the first and second events -- are presented in fig.~%
\ref{Fig3}. It is evident from this figure that the selected spectra and
time distribution are in an excellent agreement with those expected from $%
\alpha $ particles of $^{220}$Rn and $^{216}$Po. Using these results and
taking into account the efficiency of the time-amplitude analysis and the
number of accidental coincidences (3 pairs from 218 selected), the
determined activity of $^{228}$Th ($^{232}$Th family) inside the $^{116}$CdWO%
$_4$ crystals is as low as 38(3) $\mu $Bq/kg.

The same technique was applied to the sequence of $\alpha $ decays from the $%
^{235}$U family: $^{223}$Ra ($Q_\alpha $ = $5.98$ MeV, $T_{1/2}$ = $11.44$
d) $\rightarrow $ $^{219}$Rn ($Q_\alpha $ = $6.95$ MeV, $T_{1/2}$ = $3.96$
s) $\rightarrow $ $^{215}$Po ($Q_\alpha $ = $7.53$ MeV, $T_{1/2}$ = $1.78$
ms) $\rightarrow $ $^{211}$Pb. For the fast couple ($^{219}$Rn $\rightarrow $
$^{215}$Po) all events within $0.8$--$1.8$ MeV were used as triggers, while
a time interval $1$--$10$ ms (65.7\% of $^{215}$Po decays) and an energy
window 0.9--2.0 MeV were set for the second events. The obtained $\alpha $
peaks correspond to an activity of 5.5(14) $\mu $Bq/kg for the $^{227}$Ac
impurity in the crystals.

As regard the $^{226}$Ra chain ($^{238}$U family) the following sequence of $%
\beta $ and $\alpha $ decays was analyzed: $^{214}$Bi ($Q_\beta $ = $3.27$
MeV, $T_{1/2}$ = $19.9$ m) $\rightarrow $ $^{214}$Po ($Q_\alpha $ = $7.83$
MeV, $T_{1/2}\ =164.3$ $\mu $s) $\rightarrow $ $^{210}$Pb. For the first and
second events the energy threshold was equal 0.1 MeV, and a time interval of 
$100$--$1000$ $\mu $s (64.1\% of $^{214}$Po decays) was used. While the
obtained spectrum of the first pulses agrees with the model of the $\beta $
decay of $^{214}$Bi, and the distribution of the time intervals between the
first and second events can be fitted by an exponent with $T_{1/2}$ = $%
140_{-20}^{+30}$ $\mu $s (in reasonable agreement with the $^{214}$Po
half-life value), the spectrum of the second events is continuous, contrary
to the anticipated $\alpha $ peak of $^{214}$Po. Probably, part of this
continuous distribution can be explained by $^{226}$Ra contamination of the
materials neighboring the $^{116}$CdWO$_4$ crystals (optical grease, teflon,
Mylar, radon in air), while another part is caused by $^{226}$Ra decays in
the crystals. Under such an assumption activity limits for $^{226}$Ra
contaminations are derived as $\leq $0.13(3) Bq/kg for optical grease, $\leq 
$8 mBq/kg for teflon, $\leq $1.8 $\mu $Bq/dm$^2$ for Mylar, and $\leq $ 5 $%
\mu $Bq/kg for $^{116}$CdWO$_4,$ whose values do not contradict bounds
obtained earlier \cite{Dan98}. To prove these assumptions, the events
belonging to the $^{214}$Bi $\rightarrow $ $^{214}$Po $\rightarrow $ $^{210}$%
Pb chain were independently searched for in the time window of $5$--$88$ $%
\mu $s (28.9\% of $^{214}$Po decays) with the help of pulse shape analysis
(see below). For both events the energy threshold was $\approx $0.3 MeV. The
result obtained ($^{226}$Ra activity in the $^{116}$CdWO$_4$ crystals $\leq $
14 $\mu $Bq/kg) is similar to that of the time-amplitude analysis.

Finally, all couples of events found for $^{232}$Th, $^{235}$U and $^{238}$U
families as described above were eliminated from the measured data.

\subsection{Pulse-shape discrimination}

The pulse shape of the $^{116}$CdWO$_4$ scintillators in the energy region
of $0.25$--$5$ MeV is digitized by a 12 bit ADC\ and stored in 2048 channels
with 50 ns channel's width. Due to different shapes of scintillation signal
for various kinds of sources\footnote{%
It is known, that scintillation efficiency and pulse shape of inorganic
crystals depend on the local density of the energy released, hence allowing
to identify the incoming radiation (see e. g. ref. \cite{Gat62,Bart00}).} ($%
\alpha $ particles, protons, $\gamma $ quanta and cosmic muons were
investigated), the pulse-shape (PS) discrimination method based on the
optimal digital filter \cite{Gat62} was developed and clear discrimination
between $\gamma $ rays (electrons) and $\alpha $ particles was achieved \cite
{Faz98}.

The pulse shapes of enriched crystals were measured for $\alpha $ particles
with an $^{241}$Am source and for $\gamma $ rays with $^{60}$Co, $^{137}$Cs, 
$^{207}$Bi and $^{232}$Th sources in the special calibration runs\footnote{%
Because $\gamma $ rays interact with matter by mean of the energy transfer
to electrons, it was assumed that pulse shapes for electrons and $\gamma $-s
are the same. This statement was proved in the measurement with conversion
electrons of $^{207}$Bi by using the signal of the thin plastic scintillator
(placed between source and detector) as signature of the electron hitting
the $^{116}$CdWO$_4$ crystal.}. To provide an analytic description of the $%
\alpha $ or $\gamma $ signals $f_\alpha (t)$ and $f_\gamma (t)$ the pulse
shape resulting from the average of a large number of individual events has
been fitted with the sum of three (for $\alpha $ particles) or two (for $%
\gamma $-s ) exponents, giving the reference pulse shapes $\overline{f}%
_\alpha (t)$ and $\overline{f}_\gamma (t)$ (see for more details ref. \cite
{Faz98}). In the data processing the digital filter is applied to each
experimental signal $f(t)$ with aim to obtain the numerical characteristic
of its shape (shape indicator, $SI$) defined as: $SI=\sum_kf(t_k)\cdot
P(t_k-t_o)$, where the sum is over all time bins (from $k$ = 1 to $k$ =
2048), $f(t_k)$ is the digitized amplitude of a given signal (normalized to
its area) at the time $t_k$. The weight function $P(t_k-t_o)$ is determined
as $P(t)=\{\overline{f}_\alpha (t)-\overline{f}_\gamma (t)\}/\{\overline{f}%
_\alpha (t)+\overline{f}_\gamma (t)\},$ and $t_o$ is the time origin of the
signal. The measured with sources $SI$ distributions are well described by a
Gaussian functions, whose mean values and standard deviations $\sigma
_\alpha $ and $\sigma _\gamma $ have a slight energy dependence\footnote{%
For the $\gamma $-s (300--3200 keV)\quad $SI_\gamma $= 18.09 -- (4.5$\times $%
10$^{-5}E_\gamma $),\quad $\sigma _\gamma $= 2.61 -- (4.7$\times $10$%
^{-4}E_\gamma )+707/E_\gamma $, while for the $\alpha $ particles
(4000--6000 keV)\quad $SI_\alpha $= 29.0;\quad $\sigma _\alpha $= 5.11 --
(5.52$\times $10$^{-4}E_\alpha )+5520/E_\alpha $. Here all variables are
dimensionless ($E_\gamma $ and $E_\alpha $ are expressed in keV).}. For 0.9
MeV $\gamma $ quanta $SI_\gamma $ = $18\pm 3$, while for 4.8 MeV $\alpha $
particles $SI_\alpha $ = $29.0\pm 3.6$. It allows us to determine the
efficiency of the PS event selection for the different chosen intervals of $%
SI$ values ($\pm \sigma $, $\pm 2\sigma $, etc.).

\begin{figure}[tbp]
\centering
\includegraphics[width=10.cm,clip,bb=0 110 590 700] {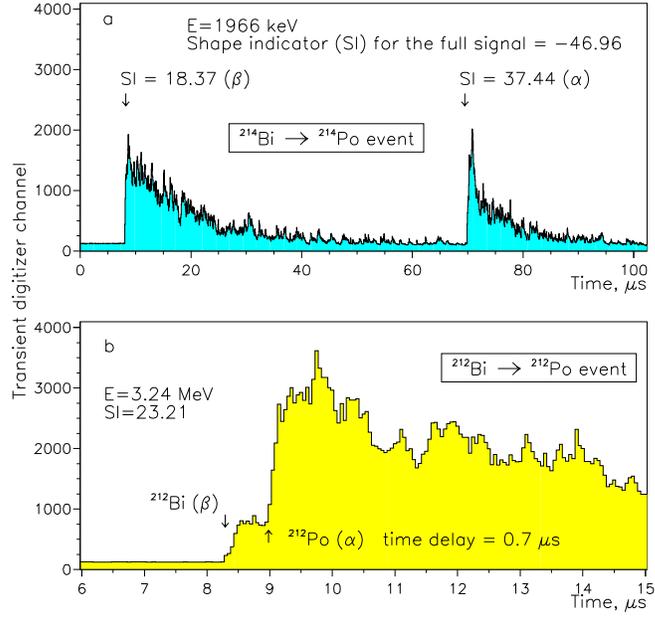}
\caption{ (a) Example of a double pulse with total energy $1.97$ MeV. The
shape indicators for the full signal and separately for its first and second
parts are $SI_\Sigma $ = $-47$;  $SI_1$ = $18.4$ (close to $SI_\gamma $) and $%
SI_2$ = $37.4$ (close to $SI_\alpha $). Most probably, this is the couple of
successive decays $^{214}$Bi ($\beta )$ $\rightarrow $ $^{214}$Po ($\alpha ;$
$T_{1/2}$ = $164.3$ $\mu $s) $\rightarrow $ $^{210}$Pb. (b) Probable event
of the chain $^{212}$Bi ($\beta )$ $\rightarrow $ $^{212}$Po ($\alpha ;$ $%
T_{1/2}$ = $0.3$ $\mu $s) $\rightarrow $ $^{208}$Pb.}
\label{Fig4}
\end{figure}

The PS selection technique ensures the very important possibility to
discriminate ''illegal'' events: double pulses, $\alpha $ events, etc., and
thus to suppress background. An example of a double pulse is shown in fig. 
\ref{Fig4}a. Value of the shape indicator for the full signal is $SI=$ --$47$%
; for the first pulse $SI_1$ = 18.4 (hence it corresponds to $\gamma $ or $%
\beta $ particle), for the second pulse $SI_2$ = 37.4 ($\alpha $ particle).
The energy release is 1.97 MeV, and without PS analysis it would be a
candidate event for $2\nu 2\beta $ decay of $^{116}$Cd.

Since the shape indicator characterizes the full signal, it is also useful
to examine the pulse front edge. For example, it was found that at least
99\% of ''pure'' $\gamma $ events (measured with calibration $^{232}$Th
source) satisfy the following restriction on pulse rise time : $\Delta t$($%
\mu $s) $\leq $ 1.24 -- 0.5$E_\gamma $ + 0.078$E_\gamma ^2$, where $E_\gamma 
$ is dimensionless variable expressed in MeV. Hence, this filter was applied
to the background data, and all events, which do not pass the test, were
excluded from the residual $\beta /\gamma $ spectrum.

\begin{figure}[tbp]
\centering
\includegraphics[width=10.cm,clip,bb=0 100 590 680] {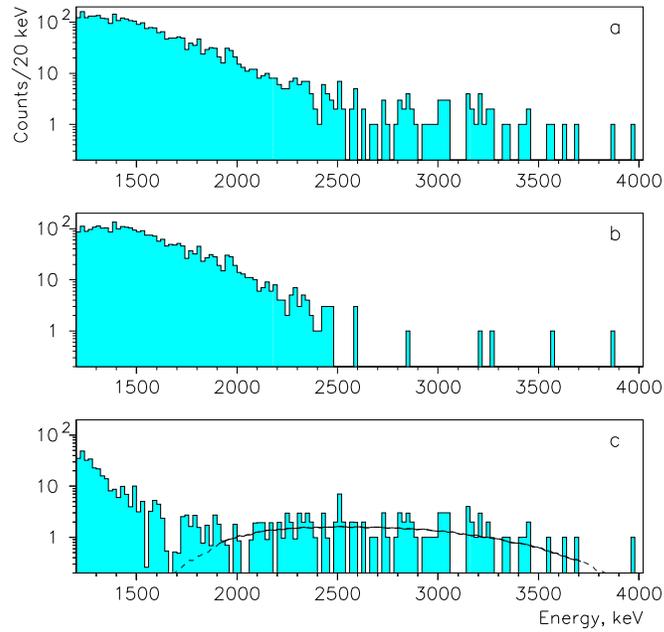}
\caption{ (a) Initial spectrum of $^{116}$CdWO$_4$ crystals (339 g, 4629 h)
in anticoincidence with shielding detectors without pulse-shape
discrimination; (b) PS\ selected $\beta /\gamma $ events (see text); (c) the
difference between spectra in fig. 5a and 5b together with the fit by the
response function for $^{212}$Bi $\rightarrow $ $^{212}$Po $\rightarrow $ $%
^{208}$Pb decay chain. The fit value is 37(4) $\mu $Bq/kg for $^{228}$Th
activity inside $^{116}$CdWO$_4$ crystals.}
\label{Fig5}
\end{figure}

\begin{figure}[tbp]
\centering
\includegraphics[width=14.cm,clip,bb=0 200 590 680] {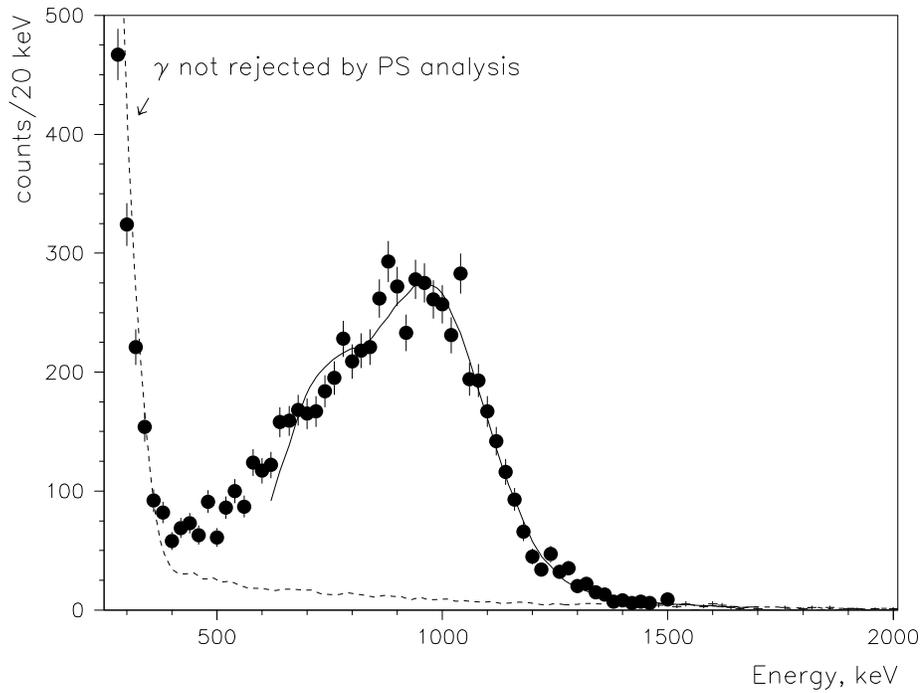}
\caption{ Spectrum after PS selection of the background events, whose $SI $
lies in the interval $SI_\gamma +2.4\sigma _\gamma <$ $SI$ $<$ $SI_\alpha
+2.4\sigma _\alpha $ (it contains $\approx $90\% of all $\alpha $ events).
The model distribution (smooth line) includes all $\alpha $-particles from
chains in $^{232}$Th and $^{238}$U families. The total $\alpha $ activity of
the $^{116}$CdWO$_4$ crystals is derived as 1.4(3) mBq/kg.}
\label{Fig6}
\end{figure}

The results of PS analysis of the data are presented in fig.~\ref{Fig5}. The
initial (without PS selection) spectrum of the $^{116}$CdWO$_4$
scintillators in the energy region $1.2$--$4$ MeV -- collected during 4629 h
in anticoincidence with active shield -- is depicted in fig.~\ref{Fig5}a,
while the spectrum after PS selection of the $\beta /\gamma $ events, whose $%
SI$ lie in the interval $SI_\gamma -3.0\sigma _\gamma \leq $ $SI\leq
SI_\gamma +2.4\sigma _\gamma $ and $\Delta t$($\mu $s) $\leq $ 1.24 -- 0.5$%
E_\gamma $ + 0.078$E_\gamma ^2$ (98\% of $\beta /\gamma $ events), is shown
in fig.~\ref{Fig5}b. From these figures the background reduction due to
pulse-shape analysis is evident. Further, fig.~\ref{Fig5}c represents the
difference between spectra in fig.~\ref{Fig5}a and \ref{Fig5}b. These
events, at least for the energy above 2 MeV, can be produced by $^{228}$Th
activity from the intrinsic contamination of the $^{116}$CdWO$_4$ crystals
(measured by the time-amplitude analysis as described above). Indeed, two
decays in the fast chain $^{212}$Bi ($Q_\beta $ = $2.25$ MeV) $\rightarrow $ 
$^{212}$Po ($Q_\alpha $ = $8.95$ MeV, $T_{1/2}$ = $0.3$ $\mu $s) $%
\rightarrow $ $^{208}$Pb can not be time resolved in the CdWO$_4$
scintillator (with an exponential decay time $\simeq $15 $\mu $s \cite{Faz98}%
) and will result in one event. The example of such an event -- recorded by
the PS acquisition system -- is depicted in fig.~\ref{Fig4}b. To determine
the residual activity of $^{228}$Th in the crystals, the response function
of $^{116}$CdWO$_4$ detectors for the $^{212}$Bi $\rightarrow $ $^{212}$Po $%
\rightarrow $ $^{208}$Pb chain was simulated with the help of GEANT3.21 code
and event generator DECAY4. The simulated function is shown in fig.~\ref
{Fig5}c, from which one can see that the high energy part of the
experimental spectrum is well reproduced ($\chi ^2$ = $1.3$) by the expected
response for $^{212}$Bi $\rightarrow ^{212}$Po $\rightarrow ^{208}$Pb decays%
\footnote{%
The rest of spectrum below 1.9 MeV (fig.~\ref{Fig5}c) can be explained as
high energy tail of the PS selected $\alpha $ particles (see fig. \ref{Fig6}%
).}. Corresponding activity of $^{228}$Th inside the $^{116}$CdWO$_4$
crystals, deduced from the fit in the 1.9--3.7 MeV energy region, is 37(4) $%
\mu $Bq/kg, that is in a good agreement with the value determined by the
time-amplitude analysis of the chain $^{220}$Rn $\rightarrow $ $^{216}$Po $%
\rightarrow $ $^{212}$Pb. Besides, the front edge analysis of 80 events with
the energy $2.0$--$4.2$ MeV ($SI\geq $ $SI_\gamma $ + 2.54$\sigma _\gamma $; 
$\Delta t$ $\geq $ 0.2 $\mu $s) was fulfilled and the half-life derived from
the average time delay between the first and second part of the signal (see
fig.~\ref{Fig4}b) is $T_{1/2}$ = $0.31(6)$ $\mu $s, in agreement with the $%
^{212}$Po table value $T_{1/2}$ = $0.299(2)$ $\mu $s \cite{Fir96}.

Fig.~\ref{Fig6} represents the spectrum after PS selection of the background
events, whose $SI$ lie in the interval $SI_\gamma +2.4\sigma _\gamma <$ $SI$ 
$<$ $SI_\alpha +2.4\sigma _\alpha $ ($\approx $90\% of $\alpha $ events).
The obtained distribution with maximum at 0.95 MeV is well reproduced by the
model, which includes all $\alpha $ particles from chains in $^{232}$Th and $%
^{238}$U families. The total $\alpha $ activity of the $^{116}$CdWO$_4$
crystals deduced from fig.~\ref{Fig6} is 1.4(3) mBq/kg. This value can be
adjusted with the activities determined by the time-amplitude analysis under
usual (for crystals) assumption that secular radioactive equilibriums in
some chains of $^{232}$Th and $^{238}$U families (like, f. e. $^{230}$Th $%
\rightarrow $ $^{226}$Ra chain) are broken.

\section{Results and discussion}

\subsection{Two-neutrino double beta decay of $^{116}$Cd}

To determine the half-life of two-neutrino 2$\beta $ decay of $^{116}$Cd,
the background in the energy interval $900$--$2900$ keV was simulated with
the help of GEANT3.21 package and event generator DECAY4. In addition to $%
^{116}$Cd two neutrino 2$\beta $ decay distribution, only three components
shown in fig.~\ref{Fig2} were used to build up the background model: $^{40}$%
K contamination of the enriched and natural CdWO$_4$ scintillators, whose
activity limits of less than 4 mBq/kg were established earlier \cite{Geo96},
and external $\gamma $ background caused by $^{232}$Th and $^{238}$U
contamination of the PMTs (one PMT for $^{116}$CdWO$_4$ crystals; one for
CdWO$_4$; two for plastic active shield)\footnote{%
The radioactive impurities of all PMTs used in the installation were
previously measured by R\&D low background set-up as (0.4--2.2) Bq/PMT and
(0.1-- 0.2) Bq/PMT for $^{226}$Ra and $^{228}$Th activity, respectively \cite
{Dan98}.}. This simple background model describes experimental data in the
chosen energy interval $900$--$2900$ keV reasonably well ($\chi ^2$ =$1.3$)
and gives the following results: the activities of $^{40}$K inside the
enriched and natural CdWO$_4$ crystals are equal 0.8(2) and 2.1(3) mBq/kg,
respectively; the half-life of two-neutrino 2$\beta $ decay of $^{116}$Cd is 
$T_{1/2}$(2$\nu $) = 2.6(1)$\times 10^{19}$ yr (only statistical
uncertainties are given, while systematical errors are pointed below).

\begin{figure}[tbp]
\centering
\includegraphics[width=14.cm,clip,bb=0 220 590 680] {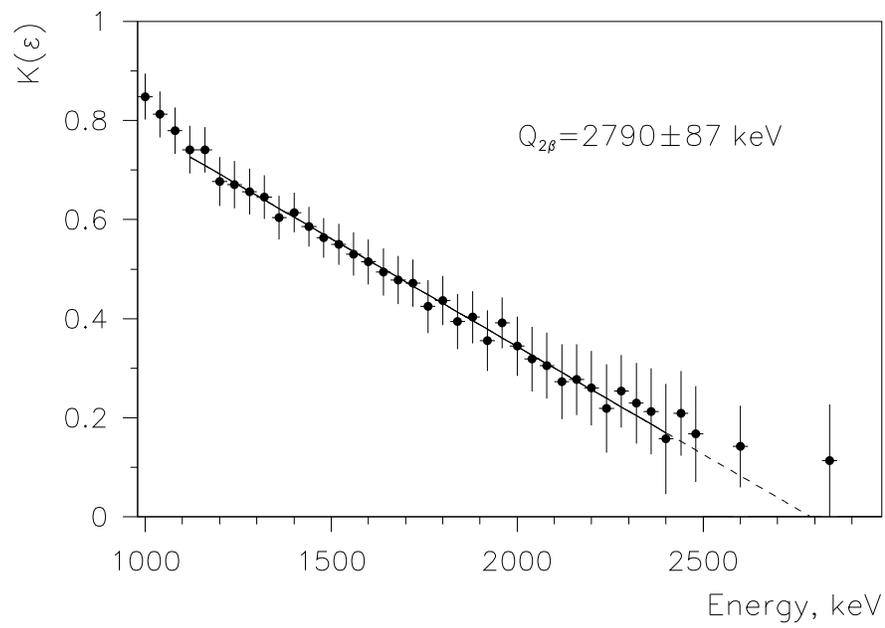}
\caption{ The $2\nu 2\beta $ decay Kurie plot and its fit by the straight
line in $1100$--$2400$ keV region. }
\label{Fig7}
\end{figure}

Taking advantage of the high statistics in our experiment (approximately
3600 events of $^{116}$Cd two neutrino 2$\beta $ decay are contained within
the interval $900$--$2900$ keV), we can prove our model with the help of
experimental $2\nu 2\beta $ decay Kurie plot: 
$K(\varepsilon )=[S(\varepsilon )/\left\{ (\varepsilon ^4+10\varepsilon
^3+40\varepsilon ^2+60\varepsilon +30)\varepsilon \right\} ]^{1/5}$, where $%
S $ is the number of events with the energy $\varepsilon $ (in electron mass
units) in the experimental spectrum after background subtraction. For the
real $2\nu 2\beta $ decay events such a Kurie plot should be the straight
line $K(\varepsilon )\sim (Q_{2\beta }-\varepsilon ),$ where $Q_{2\beta }$
is the 2$\beta $ energy release. From fig.~\ref{Fig7}, where our
experimental Kurie plot is depicted, one can see that in the region $1.1$--$%
2.4$ MeV it is well fitted by the straight line with $Q_{2\beta }=2790(87)$
keV (table value is $Q_{2\beta }=2805(4)$ keV). To take into account the
energy resolution of the detector the experimental spectrum was also fitted
by the convolution of the theoretical $2\nu 2\beta $ distribution $\rho
(\varepsilon )=A\cdot \varepsilon (\varepsilon ^4+10\varepsilon
^3+40\varepsilon ^2+60\varepsilon +30)\cdot (Q_{2\beta }-\varepsilon )^5$
with the detector resolution function (Gaussian with $FWHM$ determined as
described above) with $A$ and $Q_{2\beta }$ values as free parameters. This
fit in the energy region $1.2$--$2.8$ MeV yields a very similar value $%
Q_{2\beta }=2779(52)$ keV and an $A$ corresponding to $T_{1/2}$(2$\nu )=$
2.5(3)$\times $1$0^{19}$ yr, thus justifying our assumption that
experimental data in the region above 1.2 MeV are related mainly with $%
^{116} $Cd two neutrino $2\beta $ decay. In fact, a signal to background
ratio deduced from our data is 4:1 for the energy interval $1.2$--$2.9$ MeV,
and 15:1 for the energy range $1.6$--$2.9$ MeV, which are higher than those
reached up-to-date in other $2\beta $ decay experiments \cite{Moe94,Theo98}.

To estimate systematical uncertainties of the measured half-life, different
origins of errors were taken into account, whose contributions are listed in
Table~\ref{Tab1}. The final value is equal to:

\begin{center}
$T_{1/2}(2\nu )=2.6\pm 0.1$(stat)$_{-0.4}^{+0.7}$(syst)$\times $1$0^{19}$ yr.
\end{center}

Our result is in agreement with those measured earlier ($T_{1/2}(2\nu
)=2.6_{-0.5}^{+0.9}\times 10^{19}$ yr \cite{Eji95} and $T_{1/2}(2\nu
)=2.7_{-0.4}^{+0.5}$(stat)$_{-0.6}^{+0.9}$(syst)$\times 10^{19}$ yr \cite
{Dan95}) and disagrees to some extent with the value $T_{1/2}(2\nu )=3.75\pm
0.35$(stat)$\pm 0.21$(syst)$\times 10^{19}$ yr from ref. \cite{Arn96}%
\footnote{%
Note, that in \cite{Arn96} the quite small detection efficiency ($1.73\%$)
was calculated by the Monte Carlo method without experimental test, thus
perhaps systematical error could be higher than quoted value.}.

\subsection{New limits for 0$\nu $2$\beta $ decay of $^{116}$Cd to ground
state of $^{116}$Sn}

To estimate the half-life limit for the neutrinoless decay mode, a simple
background model was used. In fact, in the $0\nu 2\beta $ decay energy
region only three background contributions are important: (i) external $%
\gamma $ background from U/Th contamination of the PMTs; (ii) tail of the $%
2\nu 2\beta $ decay spectrum; and (iii) internal background distribution
expected from the $^{212}$Bi $\rightarrow $ $^{212}$Po $\rightarrow $ $%
^{208} $Pb decay ($^{228}$Th chain). As it was shown above, two decays in
the fast chain $^{212}$Bi $\rightarrow $ $^{212}$Po $\rightarrow $ $^{208}$%
Pb really create the background in the region of $0\nu 2\beta $ decay (see
fig.~\ref{Fig5}c). For the activity of $^{228}$Th inside the $^{116}$CdWO$_4$
crystals two values were obtained: 38(3) $\mu $Bq/kg (time-amplitude method)
and 37(4) $\mu $Bq/kg (pulse-shape analysis). Hence, in the limit of
statistical errors, we do not find an indication of a failure for the
rejection of $\alpha $ pulses by our PS analysis.

The high energy part of the experimental spectrum of the $^{116}$CdWO$_4$
crystals measured in anticoincidence with the shielding detectors and after
the time-amplitude and pulse-shape selection is shown in fig.~\ref{Fig8}.
The peak of $0\nu 2\beta $ decay is absent, thus from the data we obtain a
lower limit of the half-life: $\lim $ $T_{1/2}=\ln 2\cdot N\cdot \eta \cdot
t/\lim S,$ where $N=4$.66$\times $1$0^{23}$ is number of $^{116}$Cd nuclei, $%
t$ is the measuring time ($t=4629$ h), $\eta $ is the total detection
efficiency for $0\nu 2\beta $ decay, and $\lim S$ is the number of events in
the peak which can be excluded with a given confidence level. The value of
the detection efficiency $\eta _{MC}=0.83$ was calculated by the DECAY4 and
GEANT3.21 codes, while the efficiency of the PS analysis $\eta _{PS}=0.98$
was determined as described above, thus the total efficiency $\eta =\eta
_{MC}\cdot \eta _{PS}=0.81$. To obtain the value of $\lim S,$ the part of
the spectrum in the $1.9$--$3.8$ MeV region was fitted by the sum of the
simulated $0\nu 2\beta $ peak and three background functions (external $%
\gamma $ rays from PMT-s contamination; contribution from $^{212}$Bi $%
\rightarrow $ $^{212}$Po $\rightarrow $ $^{208}$Pb intrinsic chain; and $%
2\nu 2\beta $ decay tail). This fit gives the value of $S=-1.1\pm 1.2$
counts, which corresponds -- in accordance with Feldman-Cousins procedure
for results close to the edge of physically accepted area \cite{Feld98}
recommended by the Particle Data Group (PDG) \cite{PDG98} -- to a $\lim
S=0.9(0.3)$ counts with 90\%(68\%) C.L., and subsequently to $T_{1/2}$(0$\nu 
$2$\beta )\geq 1.5(4.6$)$\times $1$0^{23}$ yr at 90\%(68\%) C.L. However,
because of the low statistics in the energy range where the effect is
expected, the obtained values can be cross-checked in a more simple and
explicit way. Indeed, in the energy interval $2.6$--$3.1$ MeV (containing
91\% of 0$\nu $2$\beta $ peak) there is only one measured event, while the
background expected on the basis of the GEANT simulation, in the same energy
region is 3.2$_{-1.1}^{+2.1}$ counts (1.9$\pm $0.7 events from PMT
contamination; 0.4$\pm $0.1 events from 2$\nu $2$\beta $ distribution; 0.9$%
_{-0.9}^{+2}$ counts from mentioned $^{212}$Bi $\rightarrow $ $^{212}$Po $%
\rightarrow $ $^{208}$Pb chain). Following the PDG recommendation \cite
{PDG98,Feld98} we can derive from these numbers the excluded limit as $\lim
S=1.8(0.5)$ with 90\%(68\%) C.L., which leads to $T_{1/2}$(0$\nu $2$\beta
)\geq 0.7(2.5$)$\times $1$0^{23}$ yr at 90\%(68\%) C.L. confirming the
previous estimate. Finally, the following values were accepted as
conservative half-life limits for neutrinoless $2\beta $ decay of $^{116}$Cd:

\begin{center}
$T_{1/2}$(0$\nu $2$\beta )\geq 0.7(2.5$)$\times $1$0^{23}$ yr, \qquad
90\%(68\%) C.L.
\end{center}

Using calculations \cite{Sta90}, one can obtain restrictions on the neutrino
mass and right-handed admixtures in the weak interaction: $m_\nu \leq 3.0$
eV, $\eta \leq 3.$9$\times $10$^{-8}$, $\lambda \leq 3.$4$\times $1$0^{-6}$
at 90\% C.L., and neglecting right-handed contribution $m_\nu \leq 2.6(1.4)$
eV at 90\% (68\%) C.L. On the basis of calculations \cite{Arn96} we get a
similar result: $m_\nu \leq 2.4(1.3)$ eV at 90\%(68\%) C.L. In accordance
with ref. \cite{Hir96a} the value of the R-parity violating parameter of
minimal SUSY standard model is restricted by our $T_{1/2}$ limit to $%
\varepsilon \leq 8.8(6.4$)$\times $1$0^{-4}$ at 90\%(68\%) C.L.
(calculations \cite{Fae98} give more stringent restrictions: $\varepsilon
\leq 3.4(2.4$)$\times $1$0^{-4}$).

\subsection{The bounds on 0$\nu $2$\beta $ decay of $^{116}$Cd to excited
levels of $^{116}$Sn}

Not only ground state (g.s.) but also excited levels of $^{116}$Sn with $%
E_{lev}\leq Q_{2\beta }$ can be populated in 0$\nu $2$\beta $ decay of $%
^{116}$Cd. In this case one or several $\gamma $ quanta, conversion
electrons and/or e$^{+}$e$^{-}$ pairs will be emitted in a deexcitation
process, in addition to two electrons emitted in 2$\beta $ decay. The
response functions of $^{116}$CdWO$_4$ detectors for $0\nu 2\beta $ decay to
the first and second excited levels of $^{116}$Sn (2$_1^{+}$ with $%
E_{lev}=1294$ keV and 0$_1^{+}$ with $E_{lev}=1757$ keV) were simulated with
the help of GEANT3.21 and DECAY4 codes. The full absorption of all emitted
particles should result in the peak with $E=Q_{2\beta }$ (practically the
same peak as it is expected for the 0$\nu 2\beta $ decay of $^{116}$Cd to
the g.s. of $^{116}$Sn). Calculated full peak efficiencies are: $\eta
(2_1^{+})=0.14$ and $\eta (0_1^{+})=0.07$. These numbers and the value of $%
\lim S=1.8(0.5)$ with 90\%(68\%) C.L. (determined for the g.s.$\rightarrow $%
g.s. transition) give the following restrictions on half-lives of $^{116}$Cd
neutrinoless $2\beta $ decay to excited levels of $^{116}$Sn:

\begin{center}
$T_{1/2}$(g.s.$\rightarrow $ $2_1^{+})\geq 1.3(4.8$)$\times $1$0^{22}$ yr,
\qquad 90\%(68\%) C.L.,

$T_{1/2}$(g.s.$\rightarrow $ $0_1^{+})\geq 0.7(2.4$)$\times $1$0^{22}$ yr,
\qquad 90\%(68\%) C.L.
\end{center}

\subsection{Neutrinoless 2$\beta $ decay with Majoron(s) emission}

\begin{figure}[tbp]
\centering
\includegraphics[width=20.cm,clip,bb=0 200 590 680] {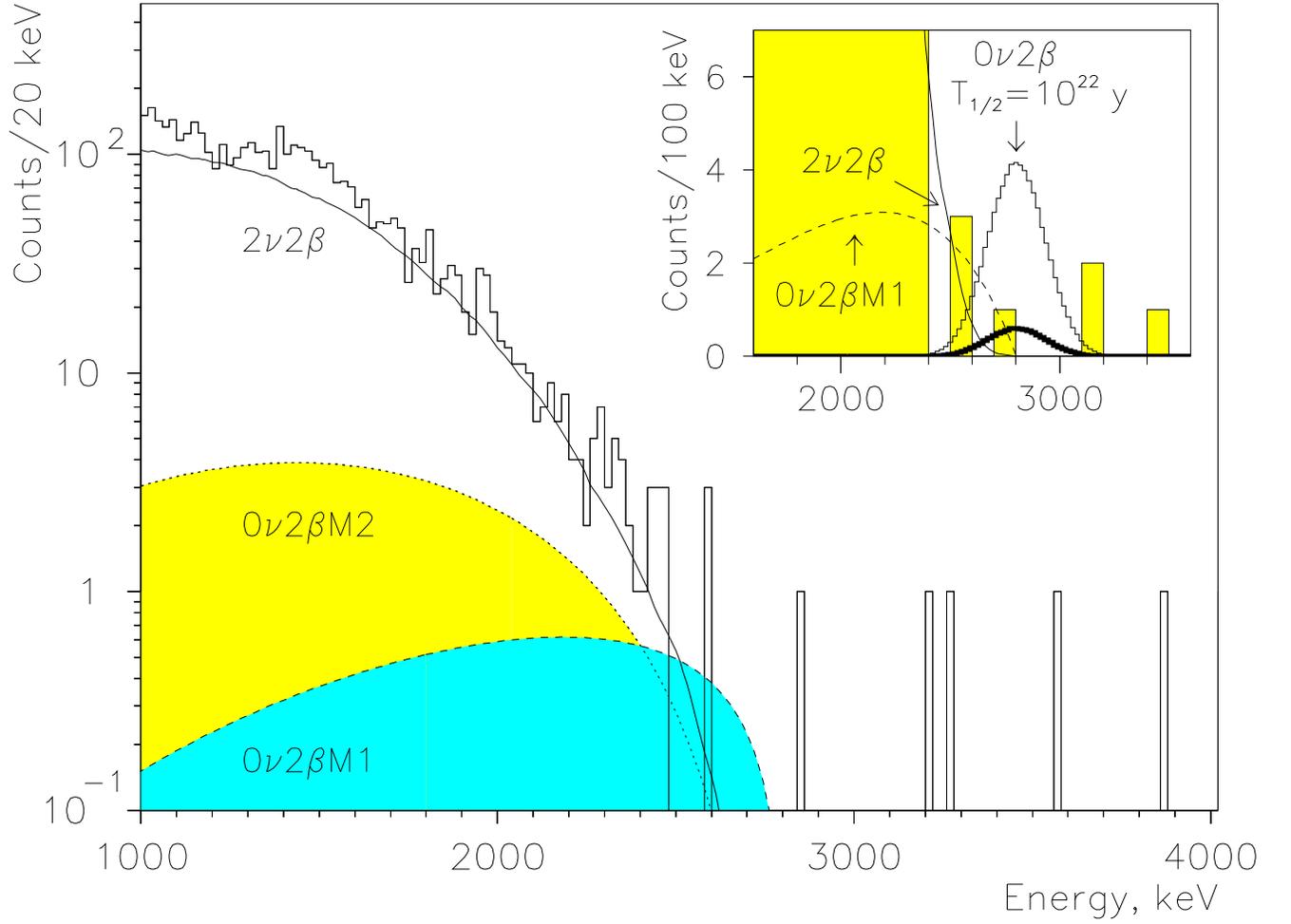}
\caption{ Part of experimental spectrum of the $^{116}$CdWO$_4$ detectors
measured during 4629 h (histogram) together with the fit from 2$\nu $2$\beta 
$ contribution ($T_{1/2}$ = 2.6$\times $1$0^{19}$ y). The smooth curves $%
0\nu 2\beta $M1 and $0\nu 2\beta $M2 are excluded with 90\% C.L.
distributions of 0$\nu $M1 and 0$\nu $M2 decay of $^{116}$Cd with $T_{1/2}$
= $3.$7$\times $1$0^{21}$ y and $T_{1/2}$ = $5.$9$\times $1$0^{20}$ y,
respectively. In the insert the expected peak from $0\nu 2\beta $ decay with 
$T_{1/2}$(0$\nu )$ = $1.$0$\times $10$^{22}$ y is shown together with the
excluded (90\% C.L.) distribution (solid histogram) with $T_{1/2}$(0$\nu )$
= $7.$0$\times $1$0^{22}$ y.}
\label{Fig8}
\end{figure}

The procedure to obtain half-life limits for 0$\nu $2$\beta $ decay with one
(two) Majoron(s) emission was carried out in two steps as follows. First,
because in the measured spectrum contributions of $^{40}$K are negligible
above the energy 1.6 MeV, the data were fitted in the energy region 1.6--2.8
MeV for 0$\nu $M1 mode (1.6--2.6 MeV for 0$\nu $M2) by using only three
theoretical distributions: $\gamma $ background from measured PMT-s
contamination ($^{226}$Ra and $^{232}$Th chains) and two neutrino 2$\beta $
decay of $^{116}$Cd, as background, and 0$\nu $2$\beta $ decay with one
(two) Majoron(s) emission, as effect. With this simple model the $\chi ^2$
value was equal to 1.1 both for 0$\nu $M1 and 0$\nu $M2 fits. As a result,
the number of events under a theoretical 0$\nu $M1 curve was determined as $%
9\pm 21$, giving no statistical evidence for the effect. It leads to an
upper limit of 41(26) events at 90\%(68\%) C.L., that together with an
efficiency value $\eta =0.905$ corresponds to the half-life limit $T_{1/2}$(0%
$\nu $M1) $\geq 3.7(5.9$)$\times $1$0^{21}$ yr. A similar procedure for 0$%
\nu $2$\beta $ decay with two Majorons emission gives $T_{1/2}$(0$\nu $M2) $%
\geq 5.9(9.4$)$\times $1$0^{20}$ yr at 90\%(68\%) C.L. The part of the
experimental spectrum and theoretical 0$\nu $M1 and 0$\nu $M2 distributions
with half-lives equal to these limits are shown in fig.~\ref{Fig8}. On one
hand, the obtained results can be treated as conservative because the
accepted background model consists of only two origins, the external
background from U/Th contamination of the PMT and 2$\nu $2$\beta $ decay
distribution, while in the energy region of interest some other sources of
background, such as the mentioned $^{212}$Bi $\rightarrow $ $^{212}$Po $%
\rightarrow $ $^{208}$Pb chain contribution from the intrinsic impurities of
the crystals, could enlarge the 0$\nu $M1 and 0$\nu $M2 limits. On the other
hand, the uncertainty of the 2$\nu $2$\beta $ decay half-life of $^{116}$Cd
could lead to the overestimated value of the 0$\nu $M1 bound. To avoid the
last possibility we have estimated the $T_{1/2}$(0$\nu $M1) limit in a more
explicit way. With this aim the energy interval 2.5--2.8 MeV -- where the
tail of 2$\nu $2$\beta $ decay distribution is practically zero (see fig.~%
\ref{Fig8}) -- was used as the most sensitive region for the 0$\nu $M1
double $\beta $ decay search. In this energy interval the number of measured
counts is 3, while the expected contribution (in the same energy range) from
the PMT contamination is 3.2$\pm $1.0 events and from 2$\nu $2$\beta $ decay
is 1.5$\pm $0.5 events, thus the total expected background is 4.7$\pm $1.1
counts. Following the PDG recommendation \cite{PDG98,Feld98} we can derive
from these numbers the excluded limit for the effect being sought as 3.0
events with 90\% C.L. Taking into account that the interval 2.5--2.8 MeV
contains 8.9\% of full 0$\nu $M1 curve, it yields a limit of $T_{1/2}$(0$\nu 
$M1) $\geq $4.5$\times $1$0^{21}$ yr (90\% C.L.), confirming the preceding
estimate:

\begin{center}
$T_{1/2}$(0$\nu $M1) $\geq 3.7(5.9$)$\times $1$0^{21}$ yr, \qquad 90\%(68\%)
C.L.,

$T_{1/2}$(0$\nu $M2) $\geq 5.9(9.4$)$\times $1$0^{20}$ yr, \qquad 90\%(68\%)
C.L.
\end{center}

Both present half-life limits are more stringent than those established in
our previous measurement during 19986 h \cite{Dan98} and in the NEMO
experiment \cite{Arn96}.

The probability of neutrinoless 2$\beta $ decay with Majoron emission can be
expressed as: $\left\{ T_{1/2}(0\nu M1)\right\} ^{-1}$= $<g_M>^2\cdot \mid $%
NME$\mid ^2\cdot G,$ where $<g_M>$ is the effective Majoron-neutrino
coupling constant, NME is the nuclear matrix element and $G$ is the
kinematical factor. Using our result $T_{1/2}$(0$\nu $M1) $\geq 3.7(5.9$)$%
\times $1$0^{21}$ yr and values of $G$ and NME calculated in the QRPA model
with proton-neutron pairing \cite{Hir96b} we obtain $g_M\leq 12(9.5$)$\times 
$1$0^{-5}$ ($g_M\leq 6$.5(5.4)$\times $1$0^{-5}$ on the basis on calculation 
\cite{Arn96}) with 90\%(68\%) C.L., which is one of the best restriction
up-to-date obtained in the direct 2$\beta $ decay experiments \cite{Moe94}.

\section{Conclusion}

The search for $^{116}$Cd double $\beta $ decay with enriched $^{116}$CdWO$%
_4 $ scintillators has entered in a new phase. The set-up with four $^{116}$%
CdWO$_4$ crystals (339 g) is running since October 1998. Improved passive
shield, new active shield made of fifteen CdWO$_4$ crystals (total mass 20.6
kg), as well as time-amplitude and pulse-shape analysis of the data result
in the reduction of the background rate in the $2.5$--$3.2$ MeV region to
0.03 counts/yr$\cdot $kg$\cdot $keV. This reduction, together with an about
threefold increase in the number of $^{116}$Cd nuclei, leads to the
substantial sensitivity enhancement of the $^{116}$Cd experiment by more
than one order of magnitude. Due to that the neutrino mass limit of $m_\nu
\leq 2.6(1.4)$ eV at 90\%(68\%) C.L. was set after the first 4629 h run.

\begin{figure}[tbp]
\centering
\includegraphics[width=10.cm,clip,bb=0 200 590 680] {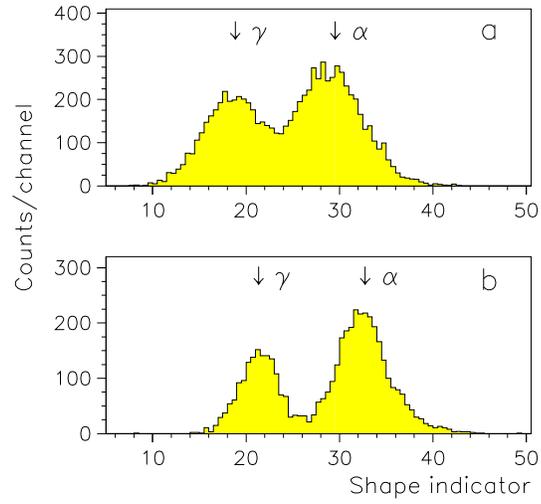}
\caption{ Shape indicator ($SI$) distributions, which represent the PS
discrimination ability of the $^{116}$CdWO$_4$ detectors: (a) for background
events (0.8--1.2 MeV) collected during 4629 h before the last upgrading; (b)
for background events (0.8--1.2 MeV) measured during 2734 h after the last
upgrading.}
\label{Fig9}
\end{figure}

In August 1999 one of our $^{116}$CdWO$_4$ crystals was annealed at high
temperature, and its light output was increased by $\approx $13\%. The PMT\
of the main $^{116}$CdWO$_4$ detectors was changed by the special low
background EMI tube (5 inches in diameter) with the RbCs photocathode, whose
spectral response better fits the CdWO$_4$ scintillation light. As a result,
the spectrometric parameters of four crystals taken as a whole were
improved. In particular, the energy resolution of the main detector is now
11.4\% at 1064 keV and 8.6\% at 2615 keV (comparing with those before this
upgrading: 14.5\% and 11\%). Besides, the PS discrimination ability of the
detector was improved too, as it is visible from fig.~\ref{Fig9}, where the $%
SI$ distribution of the measured background events -- before and after the
last upgrading -- is depicted. It is expected that after approximately 5
years of measurements the half-life limit $T_{1/2}$(0$\nu $2$\beta )\geq $ 4$%
\times $1$0^{23}$ yr will be reached which corresponds to $m_\nu \leq 1.2$
eV. The bounds on neutrinoless 2$\beta $ decay with Majorons emission and 2$%
\beta $ transitions to the excited levels of $^{116}$Sn would be improved
too.\\

\acknowledgments

The authors express their gratitude to M. Bini and O. Vihliy for their
efforts to develop and test new data acquisition system for the experiment.

\newpage
\vspace*{1.cm} 
\begin{table}[tbp]
\caption{ Different origins of the systematical uncertainties and their
contributions to the half-life value of $^{116}$Cd two neutrino $2\beta $
decay}
\begin{tabular}{|l|ll|}
\hline
Origin of the systematical error & \multicolumn{1}{|l|}{Value range} & 
Contribution to \\ 
& \multicolumn{1}{|l|}{} & $T_{1/2}(2\nu )$ value, 10$^{19}$yr \\ \hline
Live measuring time & \multicolumn{1}{|l|}{96$_{-8}^{+2}$ \%} & $+0.05,-0.2$
\\ \hline
Efficiency of PS analysis & \multicolumn{1}{|l|}{98$_{-8}^{+1}$ \%} & $%
+0.05,-0.3$ \\ \hline
Detection efficiency of $2\nu 2\beta $ decay & \multicolumn{1}{|l|}{96$\pm
4\%$} & $\pm 0.1$ \\ 
(GEANT model uncertainty) & \multicolumn{1}{|l|}{} &  \\ \hline
$^{90}$Sr--$^{90}$Y impurity in $^{116}$CdWO$_4$ & \multicolumn{1}{|l|}{$%
\leq $0.17 mBq/kg} & $+0.5$ \\ \hline
$^{234m}$Pa impurity in $^{116}$CdWO$_4$ & \multicolumn{1}{|l|}{$\leq $0.19
mBq/kg} & $+0.3$ \\ \hline
\end{tabular}
\label{Tab1}
\end{table}

\end{document}